\documentclass[10pt,conference]{IEEEtran}

\usepackage{graphicx}
\usepackage{comment}
\usepackage{booktabs}
\usepackage{float}
\usepackage{adjustbox}
\usepackage{lscape}
 \usepackage{multirow}
 \usepackage[normalem]{ulem}
 \useunder{\uline}{\ul}{}
 \usepackage{lscape}
 \usepackage{longtable}
 \usepackage{array}
 \usepackage{url}
\usepackage{algorithm}
\usepackage[noend]{algpseudocode}
\usepackage{comment}
\usepackage{float}
\usepackage{adjustbox}
\usepackage{lscape}
 \usepackage{multirow}
 \usepackage[normalem]{ulem}
 \useunder{\uline}{\ul}{}
 \usepackage{lscape}
 \usepackage{longtable}
 \usepackage{array}
\usepackage{multibib}
 \usepackage[normalem]{ulem}
 \usepackage[table,xcdraw]{xcolor}

 \useunder{\uline}{\ul}{}

\title{Supporting Contextual Conversational Agent-Based Software Development}

\author{\IEEEauthorblockN{Glaucia Melo, Luis Fernando Lins, Paulo Alencar, Donald Cowan} 
\IEEEauthorblockA{\textit{David R. Cheriton School of Computer Science} \\
\textit{University of Waterloo}\\
Waterloo, Canada \\
\{gmelo,lflinsdossantos,palencar,dcowan\}@uwaterloo.ca}
}

\begin{document}

\maketitle

\begin{abstract}

\textit 
Software Development (SD) is remarkably dynamic and is critically dependent on the knowledge acquired by the project’s software developers as the project progresses. Software developers need to understand large amounts of information related to the tasks at hand. This information (context) is often not explicit, as it can be lost in large documentation repositories, a team member’s brain, or beyond their cognitive memory capacity. These contexts include tool features, integration strategies, data structures, code syntax, approaches to tasks, project definitions, and even implicit or tacit contexts, which add significant complexity to the SD process. Current software development practices still lack sufficient techniques using the existing SD execution information and context to provide developers with relevant process guidance, augmenting their capacity to do their job using available applicable information. This paper presents ongoing and future research on an approach to support conversational agent-based knowledge-augmented software development. Developers benefit by receiving recommendations about task-related information and workflows they need to execute. This work advances human-computer interaction patterns in workflow engines, from graphical user interfaces to conversational patterns in software engineering.

\end{abstract}


\section{Introduction}
Software development (SD) is a highly complex process. Although planned initially, many changes will likely occur during an SD project’s execution \cite{MEYER2017, Murphy_Beyond2019}. SD is remarkably dynamic and is critically dependent on the knowledge acquired by the project’s software developers as the project progresses. For example, the flow of a software development process often changes during software project execution when unexpected but frequent circumstances occur \cite{Ciccio2015}. Moreover, the participants in the process, the software developers, have diverse experience and provide knowledge from different domains  with different levels of expertise \cite{gronau2005kmdl}. 

During the software development process, each participating developer uses an approach for performing their tasks highly dependent on the different characteristics of the software project and the context in which the developers operate \cite{Murphy_Beyond2019}. These contexts include tools, artifacts, integration strategies, approaches to tasks, relevant documentation, levels of expertise, project definitions, and other information (even within implicit  or tacit contexts).

Within a single company, each team might have a different context \cite{wang2007order}. For instance, one team might use agile methods while the other might use a waterfall approach. Within these teams, there are even more specific user contexts, where one team member might be a junior-level developer that uses Visual Studio Code as their IDE. In contrast, another member can be a senior-level developer that uses the Eclipse IDE. Each piece of information is considered a context or a context element, and compounding differences between multiple contexts adds significant complexity to the SD process. Recognizing context is fundamental to software engineering practices \cite{Murphy_Beyond2019,lima2015}. 

To address the complexity of working with several contexts, developers often create mental models of a project \cite{LaToza2006} by consolidating various information and context related to their specific programming tasks. For example, system details, version control, development environment, project and task specification \cite{Bradley_Fritz_Holmes_2018}, and methods of proceeding when facing that specific combination of contexts. Often, developers rely on remembering and assimilating large amounts of process-related information regarding each context combination \cite{LaToza2006}. Process-related information that varies according to the context can be easily lost or forgotten. However, developers have limited support or cognitive assistance in retrieving this knowledge. Current software development practices still lack sufficient techniques (models, designs, tools) that use the existing SD execution information and context to provide developers with relevant guidance, augmenting their capacity to do their job using available applicable information.

Many individuals and groups have examined the impact of contexts in SD. Topics include methods to support developers by managing this context \cite{Kersten_Murphy_2006}, how to provide contextual information to developers \cite{Gasparic_Murphy_Ricci_2017,Holmes_Murphy_2005}, how changes in context impact a developer’s work \cite{MEYER2017}, and how developers’ knowledge about artifacts and tasks evolve \cite{edson2020}. All approaches recognize the rich and dynamic context inherent in software development and that this context is not always explicit. SD still requires comprehensive support in terms of information and guidance based on the current context during task execution \cite{devy, Murphy_Beyond2019, MEYER2017}. Given this dynamic context, developers still lack methods based on conversational-guided agents to support the cognitive load of software development.

This paper describes a proposal and supporting research in progress that works toward realizing conversational agent-based knowledge-augmented software development solutions. Given the nature of the proposal, we believe there are three main areas to investigate, namely: (1) contexts in software engineering,
chatbots supporting software developers and (3) software process automation. We present research efforts in these main areas and critical takeaways from this research focusing on realizing a context-based chatbot solution to support developers. As specific contributions, this paper:

\begin{itemize}
\item Describes the work-in-progress to realize conversational-guided context-based agents that support software development
\item Characterizes key knowledge gained from this research
\item Proposes future research based on the current results 
\end{itemize}

This paper is organized as follows: Section \ref{section2} describes previous and ongoing research to achieve the goal of creating context-based chatbots for software development. Section \ref{futureresearch} presents future research to continue the investigation of such support tools, and Section \ref{conclusion} concludes the paper. 

\section{Context-Based Software Development Using Chatbots} \label{section2}

This section describes the work in progress to realize conversational agent-based knowledge-augmented software development solutions. First, we describe the investigation of context in software development. Then, we detail a user study and other research on chatbots to support developers. Last, we describe our efforts in integrating chatbots and workflow machines. For each, we describe the outcomes and key takeaways of our research, with the aim of exposing the findings and challenges.

\subsection{Researching Contexts in Software Engineering} \label{researchcontext}

There must be research on specific context topics in software development to realize a solution that manages the software development context. It is essential to understand the context and how to capture this context so that adaptive context recommendations about relevant software development information can be provided to developers.

Prior studies have noted the importance of the presence or absence of context information and how context recommendations and information flow during software development \cite{melo_20_knowledgereuse,Cubranic2004Learning,Gasparic2016,Ponzanelli_Bavota_Penta_Oliveto_Lanza_2014}. Some of this context relates to the information developers need to meet awareness and communication standards \cite{Ko2007,liu2021api}. Knowing how context has a significant impact on software development, we performed a study to (1) investigate the software development context elements mentioned in the literature and (2) propose a  context model \cite{melo2019context}. 

This study followed a systematic literature review (SLR) methodology \cite{Petersen2015}, which involved a structured search strategy and a backward snowballing approach \cite{jalali2012systematic} to identify relevant papers based on predefined inclusion and exclusion criteria. To ensure a well-defined focus, we adopted the Goal-Question-Metric (GQM) approach \cite{VanSolingen2002} and formulated our goals as follows:

\begin{center}
\textbf{analyze} software development 
\textbf{with the purpose of} developing a characterization 
\textbf{regarding} software developers' context 
\textbf{from the point of view of} researchers 
\textbf{in the context of} software projects
\end{center}

The aim of this SLR is to address the research question: ”What types of context have been identified by researchers in software development projects?” To provide a more comprehensive understanding of the contextual characteristics of software development, this study also seeks to answer the following nine research questions:

\begin{enumerate}
\item What are the types or classifications of context?
\item Is there a model specification technique used?
\item What are the goals or purposes of context?
\item On what step or phase of the software development does the context focus?
\item Are there any evaluations performed?
\item Are there identified limitations or gaps when using context?
\item What are the advantages or disadvantages of this context?
\item How are the context instances mined?
\item Are there any proposed abstractions?
\end{enumerate}

The results of this study demonstrate that the mentioned context elements in the literature include:
\begin{enumerate}
    \item the IDE utilized by each developer
    \item the code repository and its history
    \item organization, team, and team member activity
    \item characteristics of the person who is executing the task (including how many years of experience this person has, personal preferences)
    \item the task that is being executed, and others
\end{enumerate}

\begin{figure}[ht]
 \centering
 \includegraphics[width=0.45\textwidth]{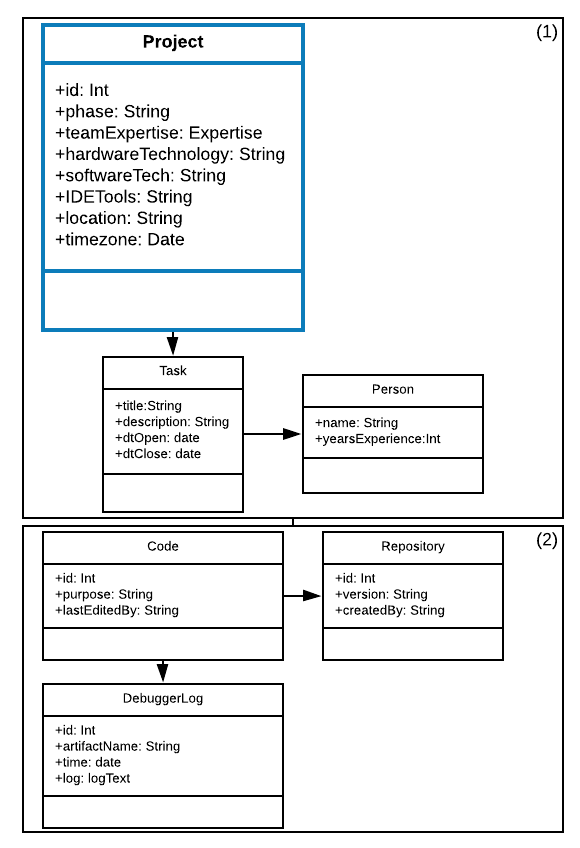}
 \caption{Extended Context Model.}
 \label{fig:contextmodelnew}
\end{figure}

After thoroughly analyzing our SLR’s most adopted elements, goals, and context representations, we developed a context model incorporating all these identified elements. This model provides a visual representation of the relationships between different context elements and highlights the intricacies of these relationships. It also serves as a framework to support any identified context and its possible variations. For instance, developers can use this model to comprehend unfamiliar contexts more easily. Figure \ref{fig:contextmodelnew} illustrates the proposed context model.

\vspace{0.2cm}
\noindent\fbox{
  \parbox{0.45\textwidth}{
  \textit{
\textbf{Key Takeaways}: Software development is surrounded by multiple context elements, such as the organization’s, the team’s, and each team member’s preferences. Our research has studied, uncovered and represented these different elements and their relationships.
  }
  }
}
\vspace{0.2cm}

\subsection{Investigating Developers' Insights on Using Chatbots to Support Their Work} \label{researchchatbots}

After investigating the context of software development, the next step in our proposal is to investigate the use of chatbots to support software development \cite{melo_20_understanding}. 

In previous work, researchers have aimed to alleviate the cognitive load of knowledge workers \cite{lima2015, biegel2015}. Developers usually work with various tools in a dynamic environment, and promoting ways to help them is critical to the quality of their work. Solutions for the issue of having to deal with multiple contexts have been proposed, but software developers are not necessarily using these solutions.

Meanwhile, a chatbot has been shown to reduce employee stress, improve productivity, and lessen training times \cite{hungerbuehler2021chatbot,casillo2021chatbot}. Given that scenario, our idea was to determine how and if a chatbot could help software developers execute their daily (heavily contextual) tasks. With that goal in mind, we performed a user study with software developers as participants. We used a within-subject methodology, where each participant was presented with a scenario describing a day in the life of a software developer, inspired by Meyer et al. \cite{MEYER2017}. Participants were asked to interact with a chatbot prototype interface (which we called DevBot) developed using Rasa and were encouraged to ask questions about the tasks with which they needed help. We collected and analyzed data from interviews, questionnaires and the  questions asked of the chatbot.

The preliminary results show that developers commonly express interest in using chatbots mainly to manage their activities (issues and tasks) and be more aware of their environment, such as the status of their colleagues, current issues assigned and processes to manage their versioning of systems. Participants have indicated that non-functional requirements such as response speed are necessary and would be essential in adopting such tools. Additionally, there are some features that the participating developers would appreciate in a chatbot that assists in software development, such as the ability to schedule meetings, support more git commands, and set priorities for tasks.

Finally, further results indicated that most developers are interested in being guided in their tasks rather than having a chatbot that can execute tasks on their behalf \cite{melo2023chatbotuserstudycomplete}. This outcome is unexpected; research and industry often pursue automation to remove the human-in-the-loop, claiming to facilitate human work.

\vspace{0.2cm}
\noindent\fbox{
  \parbox{0.45\textwidth}{
  \textit{
\textbf{Key Takeaways}: Developers are willing to use chatbots at work to be more aware of their context and to receive support at work in specific tasks; the speed of responses is important; not all developers want to have every task automated; lack of research investigating more developers' preferences.
  }
  }
}
\vspace{0.2cm}

\subsection{Integrating Processes Into Chatbots} \label{researchprocesss}

We present and discuss a novel paradigm to improve the work of software developers by providing contextual information about the tasks they are performing through cognitive and intelligent support \cite{melo_21_paradigm}. The purpose of the paradigm is to capture the current context and feed it back in a helpful way through recommendations to developers in real-time using a text-based chatbot tool as they perform the steps in the software project. By capturing context and analyzing historical data with the support of algorithms that discover patterns (machine learning), recommendations for specific tasks can be provided. These tasks are related to the software process developers are executing.

The novelty of the paradigm arises from the approaches and tools used to capture and recommend a developer’s context on the fly, considering different contexts. Applying and pursuing the proposed approach brings several advantages: less time to develop software, less effort to share knowledge among team members, enhanced collaboration, application of collective wisdom, knowledge transfer from experts to novices, and many other contributions. This research is the first where the combination of three different pillars (context, chatbots and machine learning for process navigation) have been explored to pursue appropriate recommendations during software development.

\vspace{0.2cm}
\noindent\fbox{
  \parbox{0.45\textwidth}{
  \textit{
\textbf{Key Takeaways}: The proposal in this paper is intended to stimulate thinking about the design of tools and procedures that can advance software development support by integrating context, conversational agents, process automation and machine learning.
  }
  }
}
\vspace{0.2cm}

Investigating the integration between process automation and chatbots was the next key avenue, as the feasibility of incorporating processes and contexts into chatbot domain models was still unclear. We investigated the integration of process concepts, specifically business process models, into conversational agents. This integration resulted in a new agent type, which we called process-aware conversational agents (PACAs). We believed this technique represented an initial approach toward developing a chatbot-based strategy for assisting users with everyday tasks and aiding in software development process execution. This strategy has potential applications across various process types, including software engineering \cite{Lins21_paca}.

\vspace{0.2cm}
\noindent\fbox{
  \parbox{0.45\textwidth}{
  \textit{
\textbf{Key Takeaways}: Realization of the integration between a chatbot and business processes; the demonstration of the solution using two use cases; the possibility to adapt the same solution to many other domains, including software engineering. 
  }
  }
}
\vspace{0.2cm}

\section{Envisioned Future} \label{futureresearch}

Building on the perceptions of software developers regarding context types in software engineering and process automation solutions, we aim to improve our context-based prototype design by incorporating their feedback and insights from previous context research. We are currently developing design patterns and recommendations for context-based chatbots based on our findings and supported by leading-edge research in the field. These recommendations address developers’ concerns about better integrating context with process automation in conversational agents that support their work.

Based on our findings, we have identified several actionable steps for implementing chatbot tools to support software development, including:

\begin{itemize}
    \item Contextual Support: investigate how to capture dynamic context elements related to issues and tasks, team information, and further context discovery
    \item Chatbot Features: integration of context and process into chatbots solutions, speed of response, further research to understand the intricacies of the interactions between chatbots and software developers, recommendation of process-related information
    \item Process automation: implement process automation focusing on the software development domain
\end{itemize}

\section{Conclusion} \label{conclusion}

We present an approach with ongoing and future research to support software developers in understanding their context and processes. Through recommendations based on  the task-related information they need and the process they need to execute, developers will not need to remember details of the execution of their tasks. This work advances human-computer interaction patterns in workflow engines, from graphical user interfaces to conversational patterns in software engineering.

Future work includes continuing the research to build the approach and prototype that will augment the knowledge of software developers. Several courses of action exist to realize context-based chatbot support for software developers. We intend to collect a group of identified contexts in a specific SD life cycle stage from the literature (including grey literature). The specified contexts will lead to a final context model to support knowledge-augmented SD. This model from a particular lifecycle should incorporate the contexts, levels of autonomy or controllability, and conversational patterns. Subsequently, we will incorporate this novel context model and the features developers stated in our current study into a new version of the chatbot prototype. DevBot would then comprehend the expected context, provide conversational interactions, capture an understanding of the developer, and provide the desired knowledge to developers regarding that particular SD life cycle. Finally, the prototype will be evaluated and validated qualitatively. Quantitative results can arise from comparing the efficiency of different recommendation models, considering a baseline.

Further research can expand the same approach to other software development life cycles or domains. Ultimately, we aim to pursue avenues to support developers with knowledge while motivating the importance of doing so. Simultaneously, this investigation captures essential knowledge requirements from the users who will be the tools’ consumers and software developers. The long-term research goal is to build a conversational agent-based knowledge-augmented tool that amplifies software developers’ current context awareness and knowledge.

\bibliographystyle{IEEEtran} 
\bibliography{main.bib}

\end{document}